\documentclass[preprint,showpacs,preprintnumbers,amsmath,amssymb]{revtex4}

\usepackage{graphicx}% Include figure files
\usepackage{dcolumn}% Align table columns on decimal point
\usepackage{bm}% bold math
\usepackage{color}

\newcommand{\bra}[1]{\langle #1|}
\newcommand{\ket}[1]{|#1\rangle}

\begin{document}

\title{Spin diode based on a single-wall carbon nanotube}

\author{I. Weymann}
\email{weymann@amu.edu.pl} \affiliation{Department of Physics,
Adam Mickiewicz University, 61-614 Pozna\'n, Poland}

\author{J. Barna\'s}
\affiliation{Department of Physics, Adam Mickiewicz University,
61-614 Pozna\'n, Poland} \affiliation{Institute of Molecular
Physics, Polish Academy of Sciences, 60-179 Pozna\'n, Poland}

\date{\today}

\begin{abstract}
Electronic transport through a single-wall metallic carbon
nanotube weakly coupled to one ferromagnetic and one nonmagnetic
lead is analyzed in the sequential tunneling limit. It is shown
that both the spin and charge currents flowing through such
systems are highly asymmetric with respect to the bias reversal.
As a consequence, nanotubes coupled to one nonmagnetic and one
ferromagnetic lead can be effectively used as spin diodes whose
functionality can be additionally controlled by a gate voltage.
\end{abstract}


\maketitle

Since their discovery, \cite{ijima} carbon nanotubes (CNTs) have
been extensively studied from both fundamental and application
points of view. \cite{saito98,anantram06} Owing to extreme
flexibility of nanotubes (they can be either metallic or
semiconducting), they have turned out to be ideal natural systems
to study one-dimensional electronic transport in various transport
regimes. \cite{saito98,anantram06} Transport characteristics of a
CNT contacted to metallic leads depend on the strength of
CNT-leads coupling. \cite{babic} For very good coupling, transport
reveals features typical for electron waveguides. In turn, for
weak coupling between the nanotube and leads, CNT behaves as a
large quantum dot with many orbital levels participating in
electronic transport. \cite{liangPRL02} In the intermediate case
and at sufficiently low temperatures, transport reveals features
characteristic of the Kondo phenomenon. \cite{nygard00} When the
leads are additionally ferromagnetic, transport properties of a
CNT depend on the relative orientation of the leads' magnetic
moments, leading to the so-called spin-valve effect.
\cite{tsukagoshi99,zhao02,sahoo05,manPRB06,nagabhiravaAPL06,cottetPRB06,schonenberger06,
weymannPRB07} Very recently Merchant and Markovic
\cite{markovic07} observed strong diode-like behavior in transport
characteristics of a CNT coupled to two metallic leads, one being
ferromagnetic (Co) and the second one nonmagnetic (Nb). Motivated
by this experiment, in this Letter we analyze theoretically
transport through a single-wall metallic carbon nanotube weakly
coupled to one ferromagnetic and one nonmagnetic lead, and show
that indeed such systems reveal features which are typical of spin
diodes. By considering transport in the sequential tunneling
regime, we show that both the spin and charge currents become
strongly asymmetric with respect to the bias reversal. The
magnitude and sign of this asymmetry depends on the bias voltage,
and can be additionally controlled by a gate voltage, which is of
particular interest from the application point of view. It is also
worth noting that a similar spin diode behavior has been predicted
for single semiconducting quantum dots.
\cite{rudzinskiPRB01,souzaPRB07}

To describe the spin-diode features in transport characteristics,
we consider the system consisting of a single-wall metallic CNT
which is weakly coupled to one nonmagnetic (left) and one
ferromagnetic (right) lead. Hamiltonian of the system has the
general form, $H=H_{\rm L} + H_{\rm R} + H_{\rm CNT} + H_{\rm T}$.
The first two terms describe noninteracting electrons in the
leads, $H_r = \sum_{{\mathbf k}\sigma} \varepsilon_{r{\mathbf
k}\sigma} c^{\dagger}_{r{\mathbf k}\sigma} c_{r{\mathbf
k}\sigma}$, for the left ($r={\rm L}$) and right ($r={\rm R}$)
leads, with $\varepsilon_{r{\mathbf k}\sigma}$ being the energy of
an electron with the wave vector ${\mathbf k}$ and spin $\sigma$
in the lead $r$. To describe the CNT in the weak coupling limit we
employ the model Hamiltonian introduced in Ref.~\cite{oregPRL00},
\begin{eqnarray}\label{Eq:HNT}
   H_{\rm CNT} &=& \sum_{\mu j\sigma}
   \varepsilon_{\mu j} n_{\mu j\sigma} + \frac{U}{2}
   \left( N - N_0 \right)^2
   \nonumber\\
   &+& \delta U \sum_{\mu j} n_{\mu j\uparrow} n_{\mu j\downarrow}
   + J \sum_{\mu j, \mu^\prime j^\prime}
   n_{\mu j\uparrow} n_{\mu^\prime j^\prime\downarrow}
   \,,
\end{eqnarray}
where $N=\sum_{\mu j\sigma} n_{\mu j\sigma}$, and $n_{\mu j\sigma}
= d^{\dagger}_{\mu j\sigma}d_{\mu j\sigma}$. The $j$th discrete
energy level in the subband $\mu$ ($\mu=1,2$), $\varepsilon_{\mu
j}$, is given by $\varepsilon_{\mu j} = j\Delta + (\mu-1)\delta$,
where $\Delta$ is the spacing between levels following from
quantization in a particular subband, and $\delta$ is the energy
mismatch between the level sets corresponding to the two subbands.
The charging energy of the nanotube is denoted by $U$ and $N_0$ is
the charge induced by the gate voltage. The additional Coulomb
energy of two electrons in the same level is described by $\delta
U$, while $J$ is the exchange parameter.

The tunneling Hamiltonian reads
$H_{\rm T}=\sum_{r=\rm L,R}\sum_{\mathbf k} \sum_{\mu j\sigma}(
t_{rj} c^{\dagger}_{r {\mathbf k}\sigma} d_{\mu j\sigma}+
t_{rj}^\star d^\dagger_{\mu j\sigma} c_{r {\mathbf k}\sigma})$,
where $t_{rj}$ is the tunnel matrix element between the lead $r$
and the $j$th level. Coupling of the $j$th level to external leads
can be described by $\Gamma_{rj}^{\sigma}= 2\pi |t_{rj}|^2
\rho_r^\sigma$, where $\rho_r^\sigma$ is the density of states in
the lead $r$ for spin $\sigma$. Defining the spin polarization $p$
of the ferromagnetic electrode as $p=(\rho_{\rm R}^{+}- \rho_{\rm
R}^{-})/ (\rho_{\rm R}^{+}+ \rho_{\rm R}^{-})$, one can write the
coupling of CNT to the right lead as $\Gamma_{{\rm
R}j}^{+(-)}=\alpha(1\pm p)\Gamma/2$ for the majority (minority)
electrons, while coupling to the nonmagnetic left lead as
$\Gamma^{+}_{{\rm L}j} =\Gamma^{-}_{{\rm L}j}=\Gamma/2$ (for all
levels $j$). Since CNT may be coupled to electrodes with different
coupling strengths, we introduce an asymmetry factor $\alpha$, and
assume $\alpha = 0.2$ (coupling of CNT to the ferromagnetic
electrode is weaker than to the nonmagnetic one).

When the CNT is weakly coupled to external leads, current flows
due to the sequential tunneling processes, except the Coulomb
blockade regions, where cotunneling processes dominate over the
sequential ones. However, both sequential and cotunneling
contributions to the current in the Coulomb blockade are small in
comparison to the sequential current out of the blockade regime.
The latter regime is of particular interest from the application
point of view and is relevant to many recent experiments.
Therefore, we restrict the following considerations to the
sequential transport regime, and use the real time diagrammatic
method limited to the first-order expansion. Tunneling rates in
the sequential approximation are given by the usual Fermi golden
rule. In turn, the occupation probabilities can be calculated from
the master equation $\left(\mathbf{W}\mathbf{P}\right)_{\chi}=
\Gamma\delta_{\chi\chi_0}$, \cite{thielmann,weymannPRB07} where
$\mathbf{P}$ is the vector containing the occupation
probabilities, and $\ket{\chi}$ is a many-body state of the CNT.
The elements of matrix $\mathbf{W}$ are given by $W_{\chi
\chi^\prime}=W_{\chi \chi^\prime} ^{\rm L} + W_{\chi
\chi^\prime}^{\rm R}$, with
$W_{\chi \chi^\prime}^r = 2\pi \sum_{\sigma} \rho_r^\sigma
\big\{f_r(\varepsilon_\chi - \varepsilon_{\chi^\prime})
\big|\sum_{\mu j} t_{rj}^\star \bra{\chi}d_{\mu j\sigma}^\dagger
\ket{\chi^\prime}\big|^2 + [1-f_r(\varepsilon_{\chi^\prime} -
\varepsilon_{\chi})]\big|\sum_{\mu j} t_{rj} \bra{\chi}d_{\mu
j\sigma} \ket{\chi^\prime}\big|^2 \big\}$,
for $\chi\neq\chi^\prime$, and  $W_{\chi \chi}^r = -
\sum_{\chi^\prime\neq\chi} W_{\chi^\prime\chi}^r$, where
$f_r(\varepsilon) = 1/[e^{(\varepsilon-\mu_r)/k_{\rm B}T}+1]$ and
$\mu_r$ is the electrochemical potential in the lead $r$. The
first (second) term in the bracket describes tunneling to (from)
the nanotube from (to) the lead $r$. Apart from this, one
arbitrary row $\chi_0$ of matrix $\mathbf{W}$ has been replaced by
$(\Gamma,\dots,\Gamma)$ due to the normalization condition ${\rm
Tr}\{\mathbf{P}\}=1$. The sequential current flowing through the
nanotube can be then calculated from the formula
 $ I=e/(2\hbar){\rm Tr}\{\mathbf{W}^{\rm I}\mathbf{P}\}$,
 \cite{thielmann,weymannPRB07}
where the elements of matrix $\mathbf{W}^{\rm I}$ are given by
$W_{\chi \chi^\prime}^{\rm I} = \left[
\Theta(N_{\chi^\prime}-N_{\chi}) -\Theta(N_{\chi}-N_{\chi^\prime})
\right] \left( W_{\chi \chi^\prime}^{\rm R} - W_{\chi
\chi^\prime}^{\rm L} \right)$, with $N_\chi$ being the number of
electrons in state $\ket{\chi}$ and $\Theta(x)$ denoting the
Heaviside function.

\begin{figure}[t]
  \includegraphics[width=0.75\columnwidth]{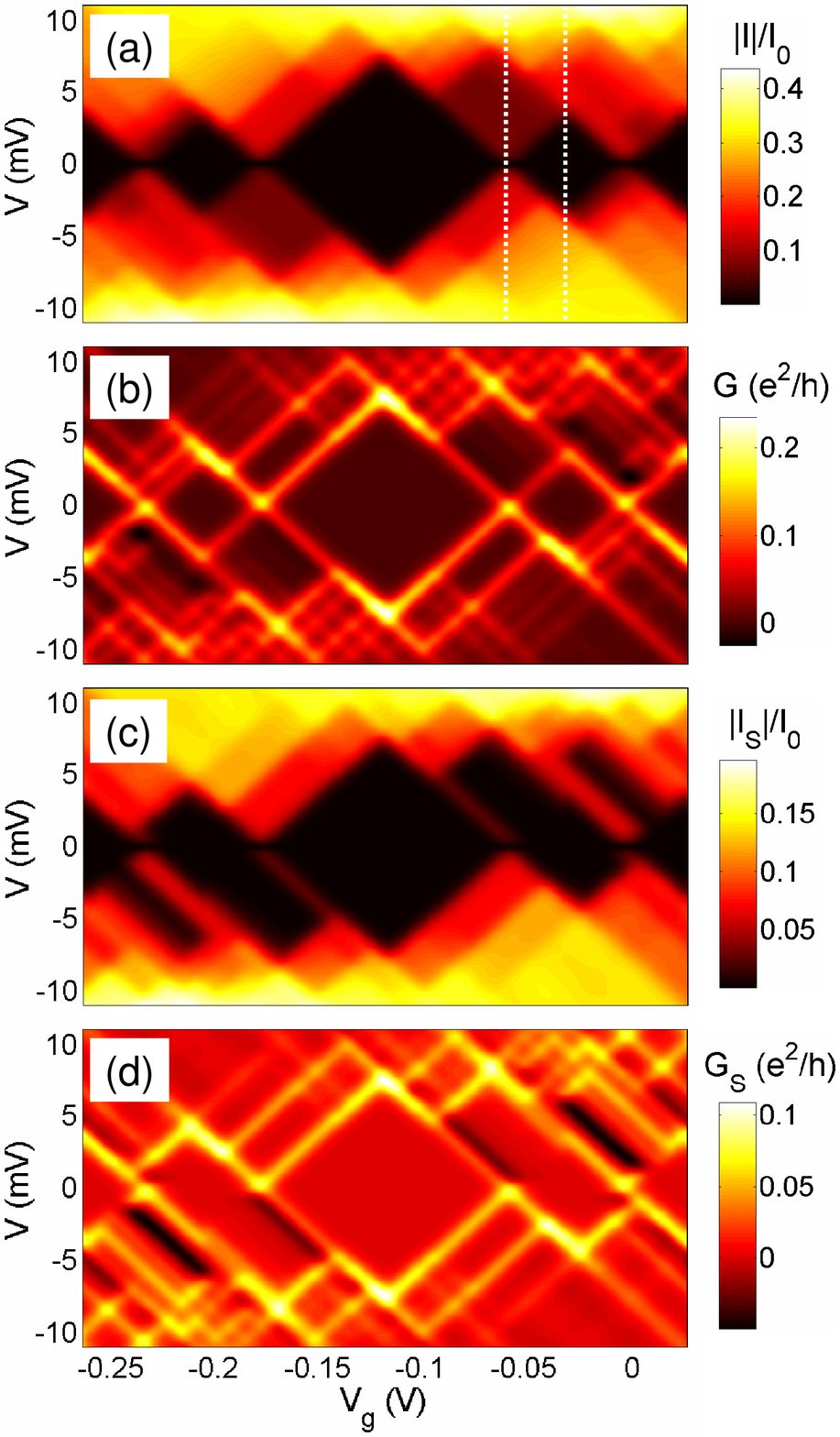}
  \caption{\label{Fig:1}
  (Color online) The absolute
  values of the current $I$ (a) and spin current
  $I_{\rm S} = I_\uparrow - I_\downarrow$ (c) in the units of
  $I_0 = e\Gamma/\hbar$,
  the differential conductance $G$ (b) and
  the differential spin conductance $G_{\rm S} = dI_{\rm S}/dV$
  (d), calculated as a function of the bias and gate voltages.
  The parameters are: $\Delta = 8.4$ meV, $U/\Delta = 0.26$,
  $J/\Delta = 0.12$, $\delta U/\Delta=0.04$,
  $\delta/\Delta = 0.27$, $k_{\rm B}T/\Delta = 0.025$,
  $p = 0.5$, $\alpha=0.2$, $x=0.14$,
  $\Gamma = 0.2$ meV, and $I_0\approx 48.7$ nA.}
\end{figure}

In order to model the single-wall metallic carbon nanotube we have
taken the parameters derived from the experiments of W. Liang {\it
et al.} \cite{liangPRL02} We have also introduced a conversion
factor, $x$, which relates the gate voltage to the electrochemical
potential shift. \cite{grabert92} In Fig.~\ref{Fig:1} we show
density plots of the absolute value of the current $I$,
differential conductance $G$, absolute value of the spin current
$I_{\rm S}$, and the differential spin conductance $G_{\rm S}$,
calculated as a function of the gate and bias voltages. The spin
current is defined as $I_{\rm S} = I_\uparrow - I_\downarrow$,
where $I_\sigma$ is the current flowing in the spin-$\sigma$
channel. $I_{\rm S}$ is related to the angular momentum current
$\tilde{I}_{\rm S}$ by the relation $\tilde{I}_{\rm S} =
(\hbar/2)I_{\rm S}/e$. Due to periodicity, in Fig.~\ref{Fig:1} we
show only one sequence of the four-fold shell filling structure of
single-wall CNTs. First of all, the Coulomb blockade regions are
clearly visible --  see the large and small black diamonds in
Fig.~\ref{Fig:1}(a). We also point on the inverse symmetry of the
transport characteristics shown in Fig.~\ref{Fig:1} with respect
to the center of the large blockade diamond.

\begin{figure}[t]
  \includegraphics[width=1\columnwidth]{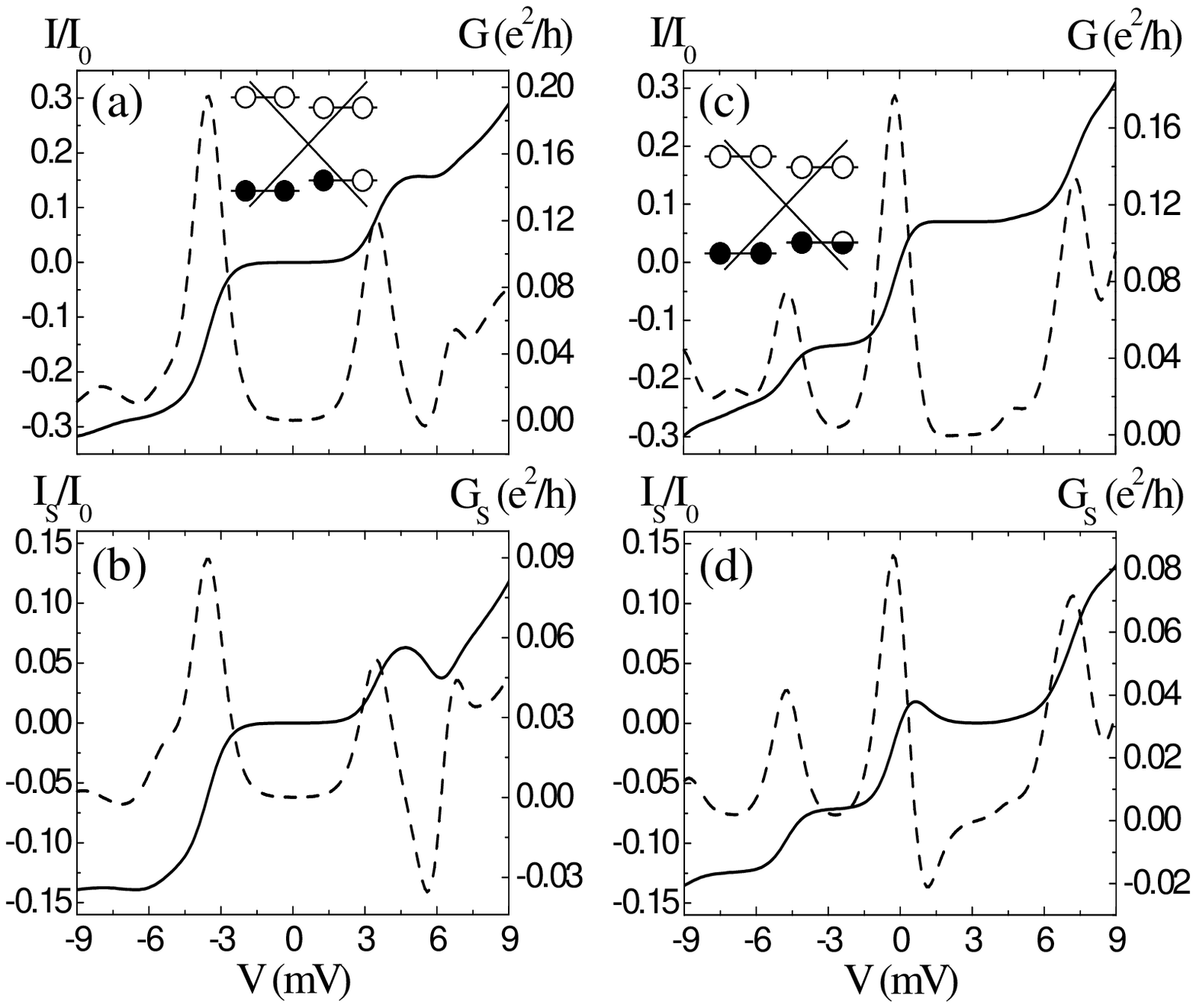}
  \caption{\label{Fig:2}
  The current (solid line) and differential conductance (dashed line) (a,c),
  and the spin current (solid line) and spin differential conductance (dashed line) (b,d)
  as a function of the bias voltage for $V_g = -0.032$ V (a,b)
  and $V_g = -0.06$ V (c,d).
  The parameters are the same as in Fig.~\ref{Fig:1}.
  The insets illustrate the two subbands of the nanotube
  with the corresponding occupations --
  the CNT is occupied with $N=3$ electrons (a)
  [empty (filled) circles correspond to empty (occupied) charge
  states], while for $V_g = -0.06$ V (c)
  the occupation number fluctuates between $N=3$ and $N=4$.}
\end{figure}

In Fig. \ref{Fig:2} we show the bias voltage dependence of the
transport characteristics (charge and spin currents and the
corresponding differential conductances) for different values of
the gate voltage, $V_g = -0.032$ V and $V_g = -0.06$ V,
respectively. These characteristics correspond to the vertical
cross-sections of Fig.~\ref{Fig:1} through the center of the first
small diamond to the right of the large one, and to the resonance
point between the large and small diamonds, respectively, see the
dotted lines in Fig.~\ref{Fig:1}(a). We recall that the large
diamond corresponds to the situation when the sequence of the four
electronic levels of the CNT becomes filled with four electrons,
while the first small blockade diamond to the right of the large
one corresponds to the situation when these four-level sequence is
occupied by three electrons only. In turn, the first small diamond
to the left of the large one describes the situation when the next
four-fold level-sequence is occupied by a single electron.
\cite{liangPRL02,weymannPRB07}

As can be seen in the figures, the current flowing through the
system is not symmetric with respect to the bias reversal. This is
associated with the asymmetry in bare tunneling matrix elements
for the spin-majority and spin-minority electrons between the CNT
and ferromagnetic lead. Due to this spin-dependence, tunneling
processes involving the spin-majority electrons are faster than
those involving the spin-minority ones. On the other hand, the
matrix elements for tunneling between the nonmagnetic lead and the
nanotube do not depend on the spin orientation. Due to the spin
dependence of tunneling processes, charge current is associated
with a nonzero spin current $I_{\rm S} = I_\uparrow -
I_\downarrow$ flowing through the system. Spin current, similarly
as the charge current, also reveals features typical of
spin-diode, see Fig. \ref{Fig:2}(b) and (d). Thus, carbon
nanotubes when coupled to one nonmagnetic and one ferromagnetic
lead can be used as spin diodes, and their functionality can be
controlled additionally by a gate voltage, see Fig. \ref{Fig:1}.

Let us now look more carefully at the two various situations shown
in Fig. \ref{Fig:2}. In Fig.~\ref{Fig:2}(a,b) we show the bias
dependence of the transport characteristics for the gate voltage
corresponding to the Coulomb blockade region with three electrons
in the last level-sequence of the CNT. We recall, that for the
parameters assumed, electrons tunnel much easier through the
barrier between the CNT and nonmagnetic lead than through the
barrier between CNT and ferromagnetic lead. At small bias the
system is in the blockade regime and current starts to flow (for
both bias polarization) when the bias voltage exceeds a threshold.
\cite{grabert92} For positive voltage electrons tunnel from the
ferromagnetic lead to the nonmagnetic one. In Fig.~\ref{Fig:2} for
$V_g = -0.032$ V the step in current for positive bias is smaller
than that for negative one. This can be explained by taking into
account spin asymmetry in tunneling matrix elements, and
difference in barriers between the left and right leads (barrier
between the magnetic leads and CNT is larger). For positive bias
transport goes mainly through the states $\ket{\uparrow\downarrow
;\uparrow}$ (lowest level doubly occupied and the next one
occupied by a spin-up electron) and $\ket{\uparrow\downarrow ;0}$
(the lowest level doubly occupied and the next one empty). In
other words, an electron leaves first the CNT and tunnels to the
nonmagnetic lead, and then a spin-up electron from the magnetic
lead tunnels to the CNT. For negative bias, in turn, transport
takes place via states $\ket{\uparrow\downarrow ;
\uparrow\downarrow }$ (both levels of the sequence are doubly
occupied) and $\ket{\uparrow\downarrow ;\downarrow}$ or
$\ket{\downarrow ;\uparrow\downarrow }$ (one of the two levels is
doubly occupied and the second one is occupied by a spin down
electron). Now, an electron tunnels first to the CNT and then a
spin up electron, either from the lower or from the higher level,
tunnels to the magnetic lead. Thus, there are now two channels for
tunneling from the CNT to the magnetic lead, which makes the step
in current at the threshold large for negative bias. For positive
bias there was only one channel open for tunneling so the current
step was smaller. When the bias increases further, additional
channel becomes open also for positive bias and the currents
become comparable. However, the two plateaus above the threshold
for positive and negative bias are significantly different. In
addition, the asymmetry in charge current leads to the
corresponding asymmetry in spin currents, see Fig.~\ref{Fig:2}(b).
Moreover, the asymmetry for spin current is even more pronounced
than for the charge current. As the charge current varies rather
monotonically with the bias voltage (positive differential
conductance), the spin current for positive bias drops with
increasing voltage above the threshold (negative differential spin
conductance). This behavior opens a functionality range for the
system as a spin diode.

Owing to the symmetry of the energy spectrum of the system, the
situation is reversed when the gate voltage admits one electron on
the last level-sequence ($V_g=-0.208$ mV in Fig.~\ref{Fig:1}). Due
to the corresponding particle-hole symmetry, transport for
positive bias goes mainly through the states $|0 ;0\rangle$ (empty
levels) and $|\uparrow ;0 \rangle$ or $|0; \uparrow \rangle$ (spin
up electron on one of the two levels). For negative bias, on the
other hand, transport takes place via states $|\uparrow\downarrow
;0\rangle$ (lower level doubly occupied) and $|\downarrow
;0\rangle$ (one spin-down electron on the lower level). Thus, now
two channels open at the threshold voltage for tunneling from the
magnetic lead to the CNT for positive bias, and therefore the
corresponding step in current is larger than for negative bias.

Consider now the situation shown in Fig.~\ref{Fig:2}(c,d). Since
the system is now at resonance, there is no current blockade and
the current increases immediately with applied voltage. Again both
the charge and spin currents for positive bias are suppressed as
compared to the currents flowing when the bias is negative. The
arguments accounting for this bias-reversal asymmetry are similar
as in the case of $V_g = -0.032$ V. It is also worth to note, that
in a broad voltage range the spin current for positive bias is
much smaller than for negative bias, see Fig.~\ref{Fig:2}(d).

We point that the spin diode behavior is observed for relatively
low transport voltages. For higher voltages there are more states
participating in transport and the current rectification is
decreased. We also notice that the operation of the spin diode can
be improved by increasing the spin polarization $p$ of the right
lead. For negative bias voltage when electrons tunnel from
nonmagnetic to ferromagnetic lead, once the nanotube becomes
occupied by a spin-down electron the current is approximately
given by $I\propto (1-p)\Gamma/2$, whereas for positive bias the
current is $I\propto \Gamma/2$. Thus, by increasing $p$, the
rectification of the current is enhanced. An ideal CNT-based spin
diode could be made by connecting the nanotube to half-metallic
lead, where $p\rightarrow 1$ due to the energy gap in one of the
two spin subbands. In addition, the operation of the diode can be
tuned by sweeping the gate voltage, see Figs.~\ref{Fig:1} and
\ref{Fig:2}.

This work, as part of the European Science Foundation EUROCORES
Programme SPINTRA, was supported by funds from the Ministry of
Science and Higher Education as a research project in years
2006-2009 and the Foundation for Polish Science.

%%%%%%%%%%%%%%%%%%%%%%%%%%%%%%%%%%%%%%%%%%%%%%%%%%%%%%%%%%%%%%%%
%%%%%%%%%%%%%%%%%%%%%%%%%%%%%%%%%%%%%%%%%%%%%%%%%%%%%%%%%%%%%%%%
%%%%%%%%%%%%%%%%%%%%%%%%%%%%%%%%%%%%%%%%%%%%%%%%%%%%%%%%%%%%%%%%

\newpage
\centerline{\large{\bf Figure captions}}

Fig. 1.
  (Color online) The absolute
  values of the current $I$ (a) and spin current
  $I_{\rm S} = I_\uparrow - I_\downarrow$ (c) in the units of
  $I_0 = e\Gamma/\hbar$,
  the differential conductance $G$ (b) and
  the differential spin conductance $G_{\rm S} = dI_{\rm S}/dV$
  (d), calculated as a function of the bias and gate voltages.
  The parameters are: $\Delta = 8.4$ meV, $U/\Delta = 0.26$,
  $J/\Delta = 0.12$, $\delta U/\Delta=0.04$,
  $\delta/\Delta = 0.27$, $k_{\rm B}T/\Delta = 0.025$,
  $p = 0.5$, $\alpha=0.2$, $x=0.14$,
  $\Gamma = 0.2$ meV, and $I_0\approx 48.7$ nA.

Fig. 2.
  The current (solid line) and differential conductance (dashed line) (a,c),
  and the spin current (solid line) and spin differential conductance (dashed line) (b,d)
  as a function of the bias voltage for $V_g = -0.032$ V (a,b)
  and $V_g = -0.06$ V (c,d).
  The parameters are the same as in Fig.~\ref{Fig:1}.
  The insets illustrate the two subbands of the nanotube
  with the corresponding occupations --
  the CNT is occupied with $N=3$ electrons (a)
  [empty (filled) circles correspond to empty (occupied) charge
  states], while for $V_g = -0.06$ V (c)
  the occupation number fluctuates between $N=3$ and $N=4$.


\newpage

\begin{thebibliography}{99}

\bibitem{ijima}
S. Iijima, Nature (London) {\bf 354}, 56 (1991).

\bibitem{saito98}
R. Saito, M. S. Dresselhaus, and G. Dresselhaus, \emph{Physical
Properties of Carbon Nanotubes} (London, UK: Imperial College
Press, 1998).

\bibitem{anantram06}
M. P. Anantram and F. Leonard, Rep. Prog. Phys \textbf{69}, 507
(2006).

\bibitem{babic}
B. Babi\'c and C. Sch\"onenberger, Phys. Rev. B {\bf 70}, 195408
(2004).

\bibitem{liangPRL02}
W. Liang, M. Bockrath, and H. Park, Phys. Rev. Lett. {\bf 88},
126801 (2002).

\bibitem{nygard00}
 J. Nyg\aa rd, D. Cobden, and P. E. Lindelof, Nature (London) {\bf 408},
342 (2000).

\bibitem{tsukagoshi99}
K.~Tsukagoshi, B. W.~Alphenaar, and H.~Ago, Nature {\bf 401}, 572
(1999).

\bibitem{zhao02}
B.~Zhao, I.~M\"onch, H.~Vinzelberg, T.~M\"uhl, and C.
M.~Schneider, Appl. Phys. Lett. {\bf 80}, 3144 (2002).

\bibitem{sahoo05}
S. Sahoo, T. Kontos, J. Furer, C. Hoffmann, M. Gr\"aber, A.
Cottet, and C. Sch\"onenberger, Nature Physics {\bf 1}, 102
(2005).

\bibitem{manPRB06}
H. T. Man, I. J. W. Wever, and A. F. Morpurgo, Phys. Rev. B {\bf
73}, 241401(R) (2006).

\bibitem{nagabhiravaAPL06}
B. Nagabhirava, T. Bansal, G. U. Sumanasekera, and B. W.
Alphenaara, L. Liu, Appl. Phys. Lett. {\bf 88}, 023503 (2006).

\bibitem{cottetPRB06}
A. Cottet and M-S. Choi, Phys. Rev. B {\bf 74}, 235316 (2006).

\bibitem{schonenberger06}
C. Sch\"onenberger, Semicond. Sci. Technol. {\bf 21}, S1 (2006).

\bibitem{weymannPRB07}
I. Weymann, J. Barna\'s, and S. Krompiewski, Phys. Rev. B {\bf
76}, 155408 (2007).

\bibitem{markovic07}
 C. A. Merchant, N. Markovic, cond-mat/0710.2297 (unpublished).

\bibitem{rudzinskiPRB01}
W. Rudzi\'nski and J. Barna\'s, Phys. Rev. B {\bf 64}, 085318
(2001); M. Wilczy\'nski, R. \'Swirkowicz, W. Rudzi\'nski, J.
Barna\'s, and V. Dugaev, J. Magn. Magn. Mater. {\bf 290-291}, 209
(2005).

\bibitem{souzaPRB07}
F. M. Souza, J. C. Egues, and A. P. Jauho, Phys. Rev. B {\bf 75},
165303 (2007).

\bibitem{oregPRL00}
Y. Oreg, K. Byczuk, and B. I. Halperin, Phys. Rev. Lett. {\bf 85},
365 (2000).

\bibitem{thielmann}
A. Thielmann, M. H. Hettler, J. K\"onig, and G. Sch\"on, Phys.
Rev. B {\bf 68}, 165341 (2003).

\bibitem{grabert92}
H. Grabert and M. H. Devoret, {\it Single Charge Tunneling}, (New
York, 1992).

\end{thebibliography}
\end{document}